\documentclass[namedreferences]{solarphysics}
\usepackage[hyperref,optionalrh,solaromanenum]{spr-sola-addons} % For Solar Physics 
\usepackage{graphicx}                    % For eps figures, newer & more powerfull
\usepackage{amssymb}                    % useful mathematical symbols
\usepackage{color}                       % For color text: \color command
\usepackage{breakurl}                         % For breaking URLs easily trough lines
\begin{document}

\begin{article}

\begin{opening}

\title{Influence of Non-Potential Coronal Magnetic Topology on Solar-Wind Models}

%%%%%%%%%%%%%%%%%%%%%%%%%%%%%%%%%%%%%%%%%%%%%%%%%%%
%% Authors Names
%
\author{S.J.~\surname{Edwards}$^{1}$\sep
       A.R.~\surname{Yeates}$^{1}$\sep
        F.-X.~\surname{Bocquet}$^{2}$ \sep
        D.H. ~\surname{Mackay}$^{3}$
       }

%%%%%%%%%%%%%%%%%%%%%%%%%%%%%%%%%%%%%%%%%%%%%%%%%%%
%% Runningheads
%
\runningauthor{Edwards \textit{et al.}}
\runningtitle{Influence of Non-Potential Coronal Magnetic Topology}

%%%%%%%%%%%%%%%%%%%%%%%%%%%%%%%%%%%%%%%%%%%%%%%%%%%
%% Affilations 
%
  \institute{$^{1}$ Department of Mathematical Sciences, Durham University, Durham, DH1 3LE, UK
                     email: \url{anthony.yeates@durham.ac.uk}\\ 
             $^{2}$ Met Office, FitzRoy Road, Exeter, EX1 3PB, UK \\
             $^{3}$ School of Mathematics and Statistics, University of St Andrews, St Andrews, KY16 9SS, UK
%                     email: \url{e.mail-c} \\
             }

%%%%%%%%%%%%%%%%%%%%%%%%%%%%%%%%%%%%%%%%%%%%%%%%%%%
%%% Abstract 
\begin{abstract}
By comparing a magneto-frictional model of the low coronal magnetic field to a potential-field source-surface model, we investigate the possible impact of non-potential magnetic structure on empirical solar-wind models. These empirical models (such as Wang--Sheeley--Arge) estimate the distribution of solar-wind speed solely from the magnetic-field structure in the low corona. Our models are computed in a domain between the solar surface and 2.5 solar radii, and are extended to 0.1 AU using a Schatten current-sheet model. The non-potential field has a more complex magnetic skeleton and quasi-separatrix structures than the potential field, leading to different sub-structure in the solar-wind speed proxies. It contains twisted magnetic structures that can perturb the separatrix surfaces traced down from the base of the heliospheric current sheet. A significant difference between the models is the greater amount of open magnetic flux in the non-potential model. Using existing empirical formulae this leads to higher predicted wind speeds for two reasons: partly because magnetic flux tubes expand less rapidly with height, but more importantly because more open field lines are further from coronal-hole boundaries.

\end{abstract}

%%%%%%%%%%%%%%%%%%%%%%%%%%%%%%%%%%%%%%%%%%%%%%%%%%%
%% Keywords
%
%\keywords{}

\end{opening}
%-------------------------------------------------

%%%%%%%%%%%%%%%%%%%%%%%%%%%%%%%%%%%%%%%%%%%%%%%%%%%
%% Sections
%
\section{Introduction}\label{s:intro} 

The structure of the magnetic field in the solar corona plays an important role in many solar phenomena, not least the interplanetary magnetic field \citep[\textit{e.g.}][]{hudson14} and the solar wind \citep[\textit{e.g.}][]{wang09}. On a global scale, the coronal magnetic topology has been studied in terms of the magnetic skeleton \citep[\textit{e.g.}][]{platten14} and also in terms of quasi-separatrix structures \citep[\textit{e.g.}][]{titov11,antiochos11}. It has been found that topological structures such as pseudostreamers and streamers can play an important role in the solar wind \citep[\textit{e.g.}][]{antiochos11,crooker12}. The network of streamers and pseudostreamers known as the ``S-web'' is thought to be connected to the acceleration of the slow solar wind via the release of plasma held in closed loops under these structures into coronal holes through magnetic reconnection \citep{antiochos11,crooker14}.

Operational forecasts by  NOAA  and the UK Met Office use the Wang--Sheeley--Arge model for the background solar wind \citep[][]{wang90, arge00}. This provides an extrapolation of the magnetic field out to $21.5\,\mathrm{R}_\odot$ (0.1 AU), along with an estimate of the radial solar-wind speed on that spherical surface. Since the model is based solely on static magnetic-field extrapolations, the speed is estimated purely from the three-dimensional magnetic-field structure, using an empirical relation (see Section \ref{s:proxies}). The topology of the low-coronal magnetic field is therefore an important element of these solar-wind forecasts. Once the magnetic-field components and speed at $21.5\,\mathrm{R}_\odot$ have been determined, these then act as the inner boundary conditions for the Enlil solar-wind model \citep[][]{odstrcil03}, which simulates the Parker spiral to calculate the solar-wind speed and field direction at 1 AU or out even further to include Mars, Jupiter, or Saturn.

The standard Wang--Sheeley--Arge magnetic field is determined in two stages: the inner stage is a potential-field source-surface (PFSS) extrapolation to $2.5\,\mathrm{R}_\odot$ from an observed boundary condition at $1\,\mathrm{R}_\odot$  \citep[][]{schatten69,altschuler69}. It assumes a purely radial magnetic field at $2.5\,\mathrm{R}_\odot$. This means that any change of polarity on the upper boundary (known at the source surface) will form a line of null points or ``null line''. This null line forms the base of the heliospheric current sheet (HCS). The second stage of the model continues the HCS out to 0.1 AU ($21.5\,\mathrm{R}_\odot$). To do this, a second potential field is generated between $2.5\,\mathrm{R}_\odot$ and $21.5\,\mathrm{R}_\odot$, whose lower boundary condition is the absolute value of radial magnetic field ($|B_r|$) on the source surface of the inner PFSS extrapolation.  A potential field is then extrapolated assuming that it decays to zero at infinity. Once this field is generated, it is reversed where it connects to a patch of field that was negative in the inner part, thus creating infinitesimally thin current sheets between oppositely directed fields. This is known as the Schatten current sheet model \citep[][]{schatten71}. It is implemented primarily to smooth out latitudinal gradients in $|\mathbfit{B}|$, producing a latitudinally uniform heliospheric magnetic field in closer accordance with \textit{Ulysses} observations \citep{smith01}.

The aim of this article is to study the effect of replacing the innermost PFSS extrapolation in the WSA model (up to $2.5\,\mathrm{R}_\odot$) with a more sophisticated, non-potential, magnetic-field model. We focus on the qualitative differences that may be expected when the potential-field assumption is removed, as preparation for future validation using the Enlil model. The work is motivated by significant shortcomings of the potential-field assumption, which does not allow electric currents to form. Yet currents are manifestly present in the low corona: not only do we observe twisted magnetic structures directly, but the associated free energy required to power flares or coronal mass ejections is (by definition) not present in potential fields.

Recent years have seen the development of several non-potential models for the global coronal magnetic field \citep[see][for a review]{mackay12}, ranging from full-MHD models including plasma thermodynamics \citep[\textit{e.g.}][]{riley11} to static nonlinear force-free field extrapolations \citep[\textit{e.g.}][]{tadesse14}. 

\citet{riley06} compared a full MHD model with the corresponding PFSS model for four Carrington rotations during Solar Cycle 23. They found that many features were similar between the two approaches, such as the boundaries of the coronal holes, although less open flux was found in the PFSS models. Notable differences were found in the heights reached by closed coronal loops, suggesting that the ``source surface" used in the PFSS model should not be spherical.

An alternative, but less computationally expensive, model known as the current sheet source surface (CSSS) model has been presented by \citet{zhao95}. This sets a cusp surface (normally at $2.5\,\mathrm{R}_\odot$) and an outer source surface (normally $15\,\mathrm{R}_\odot$) where the field lines are forced to be radial. The model is based on a magnetohydrostatic solution that includes large-scale horizontal currents, in addition to sheet currents above the cusp surface (similar to the Schatten current-sheet model). This model has been validated against the PFSS model and also against solar-wind observations \citep{poduval14} and has been found to provide better correlation with in situ solar-wind observations than the PFSS model.

Here, we compare PFSS extrapolations with the magneto-frictional model, in which electric currents and free magnetic energy are built up quasi-statically as the coronal magnetic field is sheared by surface-footpoint motions \citep{vanballegooijen00}. The model is sufficiently simple to simulate continuously months to years of coronal evolution, but sufficiently detailed to include the time-dependent build up of electric currents, magnetic helicity, and free magnetic energy. The resulting magnetic topology is substantially different from PFSS extrapolations, with the formation of twisted magnetic flux ropes and very different magnetic connectivities. It was previously shown by \citet{yeates10} that the presence of current causes magnetic-field structures to expand and so influences the amount of open field present. Near solar maximum the effect is particularly important, since the global magnetic field is dominated by active region fields, which can be highly non-potential. Thus we expect a significant difference in the predicted solar wind compared to the PFSS-based WSA model, as we will demonstrate in this article.

\section{Coronal Magnetic Field Models}\label{s:models}

In this article, we compare a potential-field source-surface (PFSS) model and a non-potential (NP) model based on the magneto-frictional method for two dates: 4 April 2000 and 30 April 2013. Figure \ref{fig:3D_fieldlines} shows that the two models lead to different three-dimensional magnetic-field structures, which we will analyse in Section \ref{s:top}. Both  dates are taken from solar maximum, the first from the Cycle 23 maximum and the second from the Cycle 24 maximum. We chose solar maximum dates as these are expected to show more significant differences between the two models, as well as being the phase when the solar-wind structure is least well understood.

We have used identical boundary conditions for the photospheric radial magnetic field in the PFSS and NP models. The NP model for a particular day must be generated by evolution over a sufficient integration time, during which the evolution of $B_r$ on the full solar photosphere is required. Accordingly, the NP model has been evolved in tandem with a surface flux-transport simulation for the photospheric radial field \citep[\textit{e.g.}][]{sheeley05, mackay12}, as described by \citet{yeates14}. The simulation was initiated on 15 June 1996 and new bipolar magnetic regions assimilated during the evolution, based on US National Solar Observatory/Kitt Peak and SOLIS (Synoptic Optical Long-term Investigations of the Sun) data\footnote{\url{http://solis.nso.edu/vsm/vsm_maps.php}}. This integration time is sufficient to allow reasonable electric currents to build-up in the NP model in a self-consistent way. The minimum integration time necessary for i) the total coronal current and ii) the open magnetic flux to reach steady levels is only about two months, although \citet{yeates12} found that the topology of high-latitude magnetic structures can have a memory of two years or more in the NP model.

\subsection{Potential-Field Source-Surface Model}
A potential field is extrapolated from the boundary condition at $1\,\mathrm{R}_\odot$ where the radial component of the magnetic field is taken from the flux transport model. Additionally the assumption is made that at $2.5\,\mathrm{R}_\odot$ the field is purely radial. The potential field is found by solving Laplace's equation in spherical coordinates and the solution is given in terms of spherical harmonics. Ideally, an infinite number of harmonics should be included; however, for numerical reasons it is necessary to truncate this sum and in this case we sum up to a maximum harmonic number $l_{\mathrm{max}}=85$. This model is static and is extrapolated from the simulated photospheric magnetic field on a particular day.

\subsection{Magneto-frictional Model}

The model is described in more detail by \citet{yeates14}. In contrast to static potential field extrapolations, the magneto-frictional model follows the continuous time-evolution of the coronal magnetic field, driven by the photospheric evolution. The developing magnetic field structure is significantly more complex than the potential field, with large-scale electric currents both in active regions and in the quiet Sun \citep{yeates2008c}.

The time evolution of the magnetic field is found by solving the uncurled induction equation for the vector potential [$\mathbfit{A}$],
\begin{equation}
\frac{\partial \mathbfit{A}}{\partial t}=\mathbfit{v} \times \mathbfit{B}-\mathbf{\mathcal{E}},
\label{eq:induction}
\end{equation}
where $\mathbfit{B}=\nabla \times \mathbfit{A}$. Ohmic diffusion is neglected but we include a hyperdiffusion of the form
\begin{equation}
\mathbf{\mathcal{E}}=-\frac{\mathbfit{B}}{B^2}\nabla \cdot (\eta_4B^2\nabla \alpha),
\end{equation}
where
\begin{equation}
\alpha = \frac{\mathbfit{B} \cdot \mathbfit{j}}{B^2}
\end{equation}
is the current helicity density, $\mathbfit{j}=\nabla \times \mathbfit{B}$ is the current density and $\eta_4=10^{11}$ km$^4$ s$^{-1}$ \citep[see][]{vanballegooijen08}. The hyperdiffusion simulates the mean effect of small-scale turbulence in the coronal magnetic field, allowing magnetic reconnection but preserving magnetic helicity in the volume.

The velocity is determined using the magneto-frictional technique \citep{yang86, craig86} and is given by
\begin{equation}
\mathbfit{v}=\frac{1}{\nu}\frac{\mathbfit{j}\times \mathbfit{B}}{B^2}+v_{\mathrm{out}}(r)\mathbfit{e}_r.
\end{equation}
The first term causes the system to relax to a force-free equilibrium, while the second term is a radial outflow imposed near $r=2.5\,\mathrm{R}_\odot$ to ensure that the field is approximately radial there, whilst allowing horizontal magnetic structures such as flux ropes to be ejected through the boundary. It is important to note that this outflow velocity is uniform in $(\theta,\phi)$ and is imposed, so it cannot be used to model the spatial and temporal distribution of solar wind speed. Rather, in Section \ref{s:proxies}, we consider an empirical model of the solar wind speed based only on the magnetic structure, as in the WSA model.

\begin{figure}
\centering
\includegraphics[width=\linewidth]{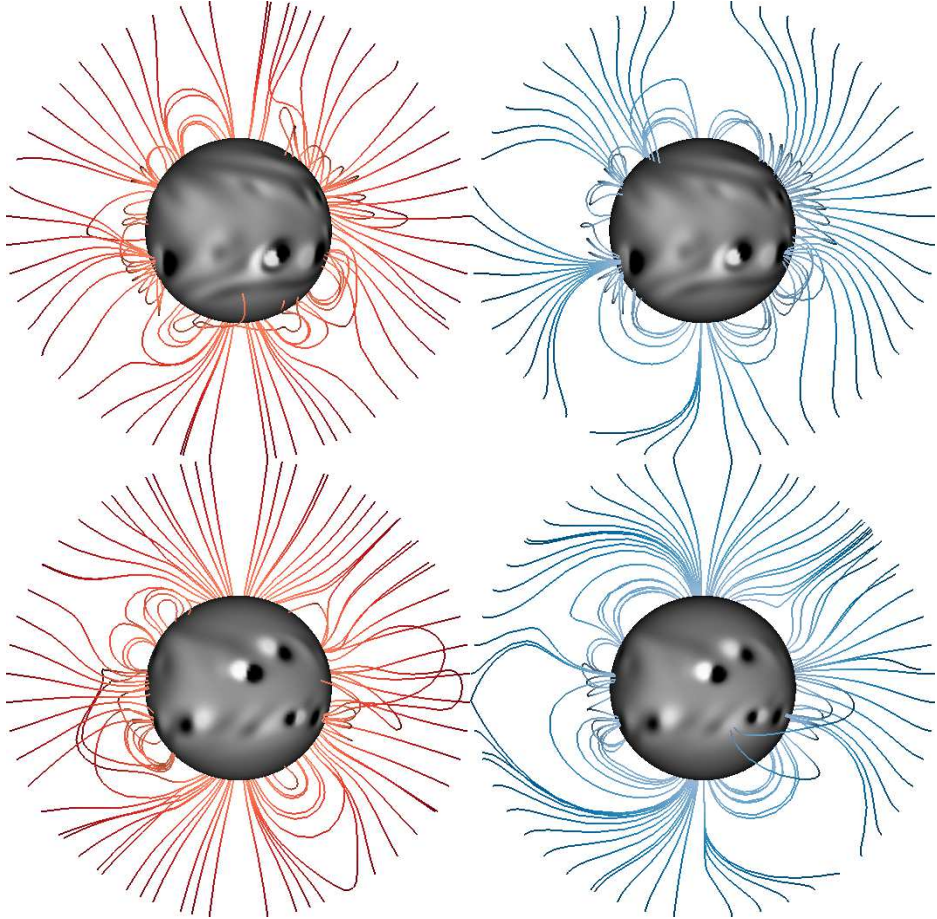}
\caption{Magnetic field lines for 4 April 2000 (top row) and 30 April 2013 (bottow row) for the NP model (left column) and PFSS model (right column). The solar surface is shaded with $B_r(R_\odot)$ (white positive, black negative).}
\label{fig:3D_fieldlines}
\end{figure}

\subsection{Schatten Current-Sheet Model}
As mentioned earlier, in the forecasting of the solar wind it is necessary to know how the magnetic field behaves between the $2.5\,\mathrm{R}_\odot$ source surface and the 0.1 AU inner boundary of the Enlil code. In the volume between $2.5\,\mathrm{R}_\odot$ and $21.5\,\mathrm{R}_\odot$ a potential field is extrapolated outward from the absolute value $|B_r|$ at $2.5\,\mathrm{R}_\odot$, assuming all magnetic-field components decay to zero at infinity. The original field direction is then restored along field lines originating in regions of negative $B_r$ on the $2.5\,\mathrm{R}_\odot$ source surface. This produces infinitesimally thin current sheets and is known as the Schatten current-sheet model \citep{schatten71}. The main advantage of this model is that it spreads the distribution of magnetic flux across the outer boundary more evenly than that found from a simple radial extrapolation of the field.

We implement the Schatten current-sheet model in the same way for both the PFSS and NP models, using their respective distributions of $B_r$ on the source surface at $2.5\,\mathrm{R}_\odot$. Note that our model contains only sheet currents in the outer region, unlike the Current Sheet Source Surface model of \citet{zhao95}, which also includes horizontal volume currents, controlled by a single additional free parameter. In principle the latter may allow for improved agreement with solar-wind observations if the part of this model below the cusp surface ($2.5\,\mathrm{R}_\odot$) was replaced by the NP model, but here we simply employ both of our lower coronal models in conjunction with the original Schatten model.

\begin{figure}
\centering
\includegraphics[width=\linewidth]{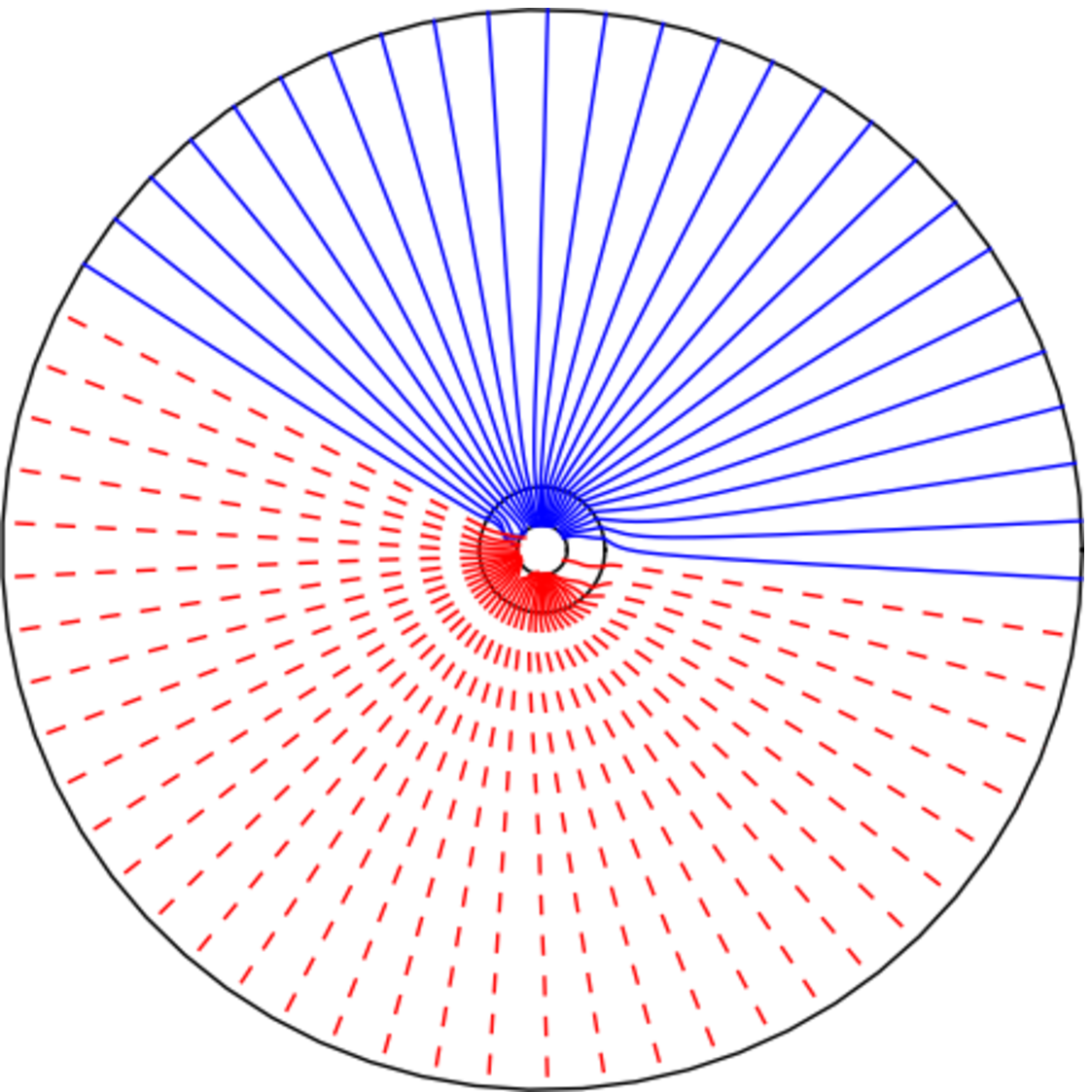}
\caption{Magnetic-field lines traced in the plane of sky for the combined NP and Schatten current-sheet model for 30 April 2013. Red-dashed lines show field directed out of the Sun, blue-solid lines show field directed into the Sun. The NP model is in the inner shell, while the current sheet model is in the much larger outer shell. }
\label{fig:scs_pos}
\end{figure}

Figure \ref{fig:scs_pos} shows some field lines traced in the plane of sky for the combined NP and Schatten current-sheet model for 30 April 2013. There are infinitesimally thin current sheets between the inwardly directed (blue solid) field lines and the outwardly directed (red dashed) field lines. In the Schatten current-sheet part of the model, the field lines are mostly radial except near to the $2.5\,\mathrm{R}_\odot$ boundary where they spread out around streamer and pseudostreamer structures.

\section{Effect on Magnetic Topology} \label{s:top}
We investigate the magnetic topology in several ways. Firstly we consider the regions of open field on the photosphere by tracing field lines. This will highlight any differences in the sizes and shapes of the footpoints of open-field regions. Secondly, we consider the magnetic skeleton \citep[\textit{e.g.}][]{longcope05}, which is a network of null points and their associated separatrix structures that divide space into topologically distinct flux domains. In addition, we look for regions where field lines are highly divergent. This is often known as the quasi-skeleton and is measured by a quantity known as the ``squashing factor" or $Q$ \citep{titov02}.

\subsection{Open-Field Regions}
\begin{figure}
\centering
\includegraphics[width=\linewidth]{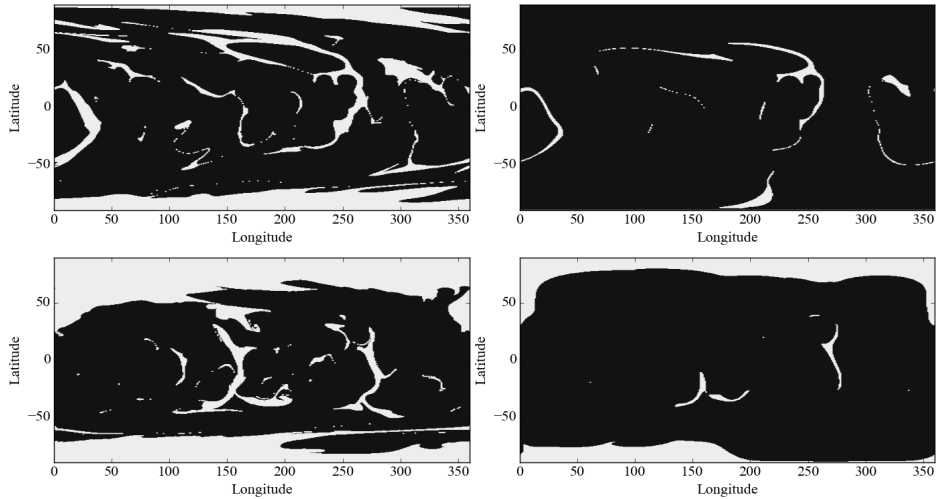}
\caption{Map of open magnetic field regions on the photosphere for 4 April 2000 (top) and 30 April 2013 (bottom) in the NP field model (left) and the PFSS model (right). Black indicates closed field line footpoints, while white indicates open field line footpoints.}
\label{fig:coronal_holes}
\end{figure}

The difference in size and shape of photospheric open-field regions between the two models is evident in Figure \ref{fig:coronal_holes}, which maps the location of coronal-hole footpoints on the photospheric boundary for each case. 
We can see that on both dates in the PFSS extrapolations (right column) the coronal holes are much smaller than in the NP model (left column). Also the polar coronal holes are not present in the PFSS model for 4 April 2000 (top right) whereas they are present in the NP model for the same date (top left). Otherwise, we can see that some coronal holes are seen in similar positions and with similar shapes in the PFSS and NP models; this is to be expected since they have a common lower boundary condition.

\subsection{Magnetic Skeleton}

In three dimensions the field lines that pass through a null point form two structures: a two-dimensional separatrix surface and a one-dimensional spine line \citep[\textit{e.g.}][]{priest96,parnell96}. When two separatrix surfaces intersect they form a separator. In global models, additional features of the magnetic skeleton are HCS curtains. These are special separatrix surfaces that are traced down from the null line on the $2.5\,\mathrm{R}_\odot$ boundary in PFSS models \citep[see also][]{platten14}.

We find the magnetic nulls using the trilinear method of \citet{haynes07}. From these null points we trace out the magnetic skeleton of separatrix surfaces using the method described by \citet{haynes10}. This method works by taking a ring of points in the fan plane of the null close to the null point, and mapping these points out along the field line until they hit either another null or the boundary. If they hit another null a separator is traced back to the initial null and the ring is broken and the points mapped along the spine. 

The null point finding method \citep[\textit{e.g.}][]{cook09,platten14,freed15,edwards15a} and similarly the skeleton finding method \citep[\textit{e.g.}][]{vandriel12,platten14,edwards15b} have both been used to find the global coronal topology in PFSS extrapolations. We now expand on this work by applying these same methods in the global NP model of the solar corona. 

\begin{figure}
\centering
\includegraphics[width=\linewidth]{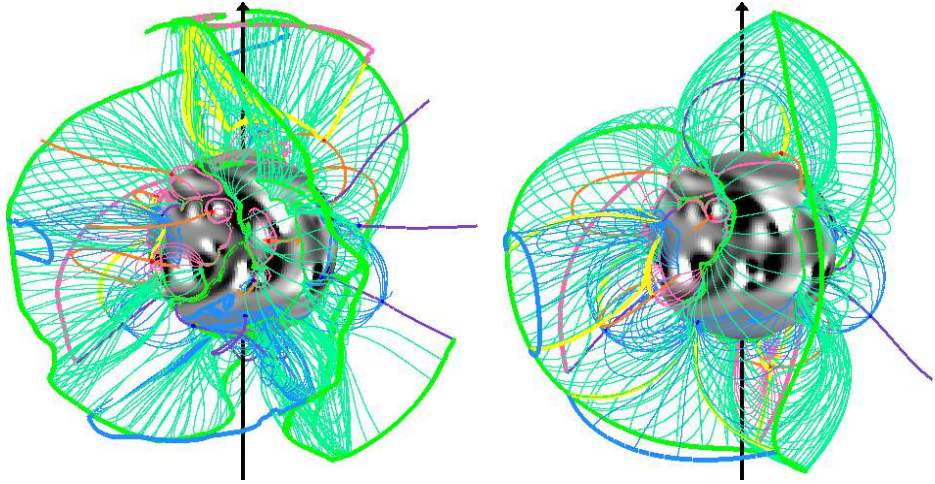}
\caption{Three-dimensional representation of the magnetic skeleton between $1\,\mathrm{R}_\odot$ and $2.5\,\mathrm{R}_\odot$ for 4 April 2000. Left shows the NP model and right the PFSS model. Red and Blue dots represent positive and negative null points, respectively. Thin blue, pink, and green lines represent field lines in the separatrix surfaces from negative nulls, positive nulls, and the HCS null line respectively. Yellow lines represent separators. Purple and orange lines represent the spines of negative and positive nulls, respectively. The solar surface is shaded with $B_r(R_\odot)$ (white positive, black negative). (Animated versions of these figures are available in the Electronic Supplementary Materials.)}
\label{fig:skeleton3D_00100}
\end{figure}
\begin{figure}
\includegraphics[width=\linewidth]{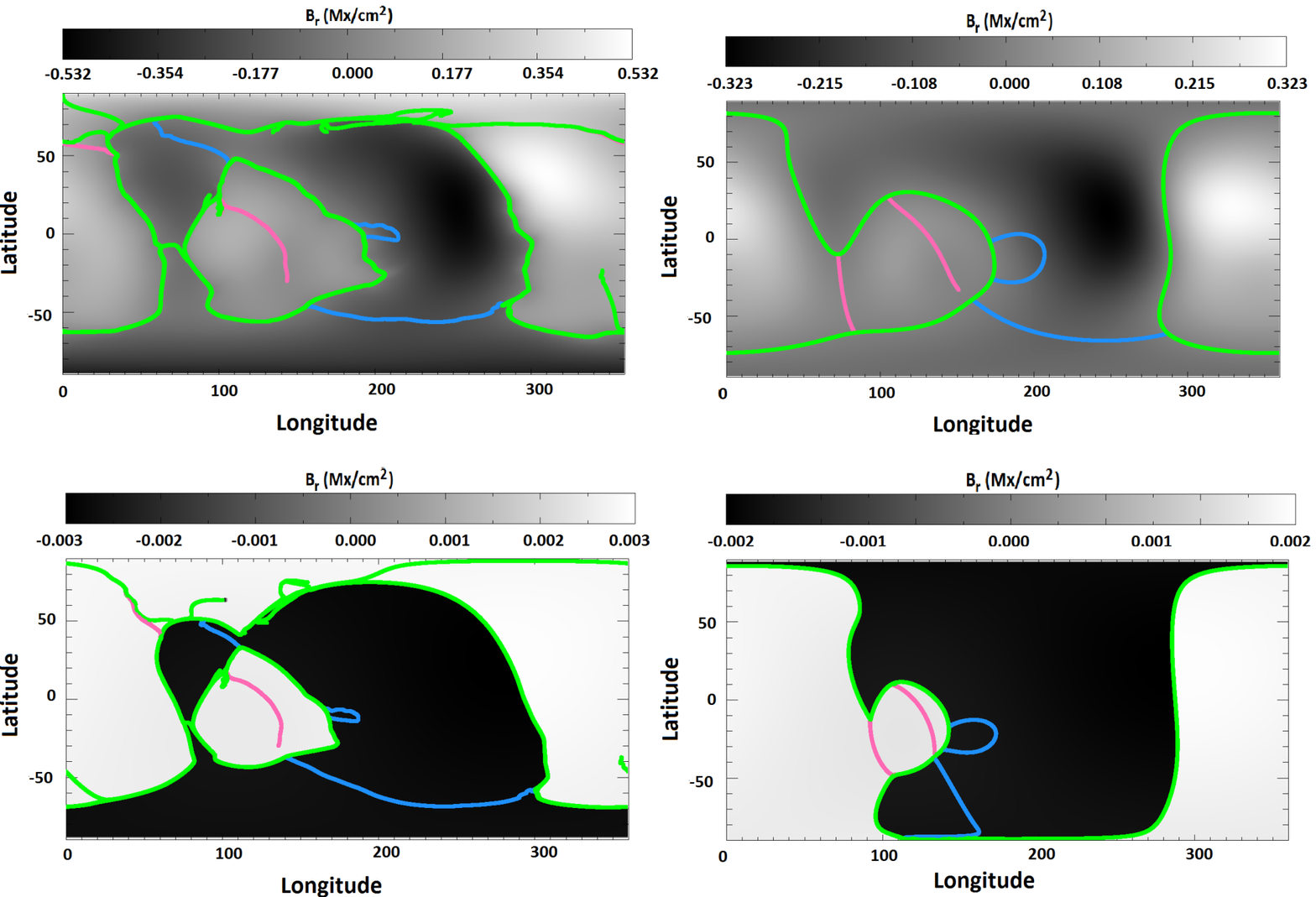}
\caption{Cuts through the magnetic skeleton at $2.5\,\mathrm{R}_\odot$ (top) and $21.5\,\mathrm{R}_\odot$ (bottom) for the NP model (left) and PFSS model (right) for 4 April 2000. The green lines represent the intersection of the HCS with the surface; the pink and blue lines represent the intersection of separatrix surfaces from positive and negative nulls, respectively, with the surface. The background is shaded with $B_r$ (white positive, black negative).}
\label{fig:sepsurf_cuts_00100}
\end{figure}

\begin{figure}
\centering
\includegraphics[width=\linewidth]{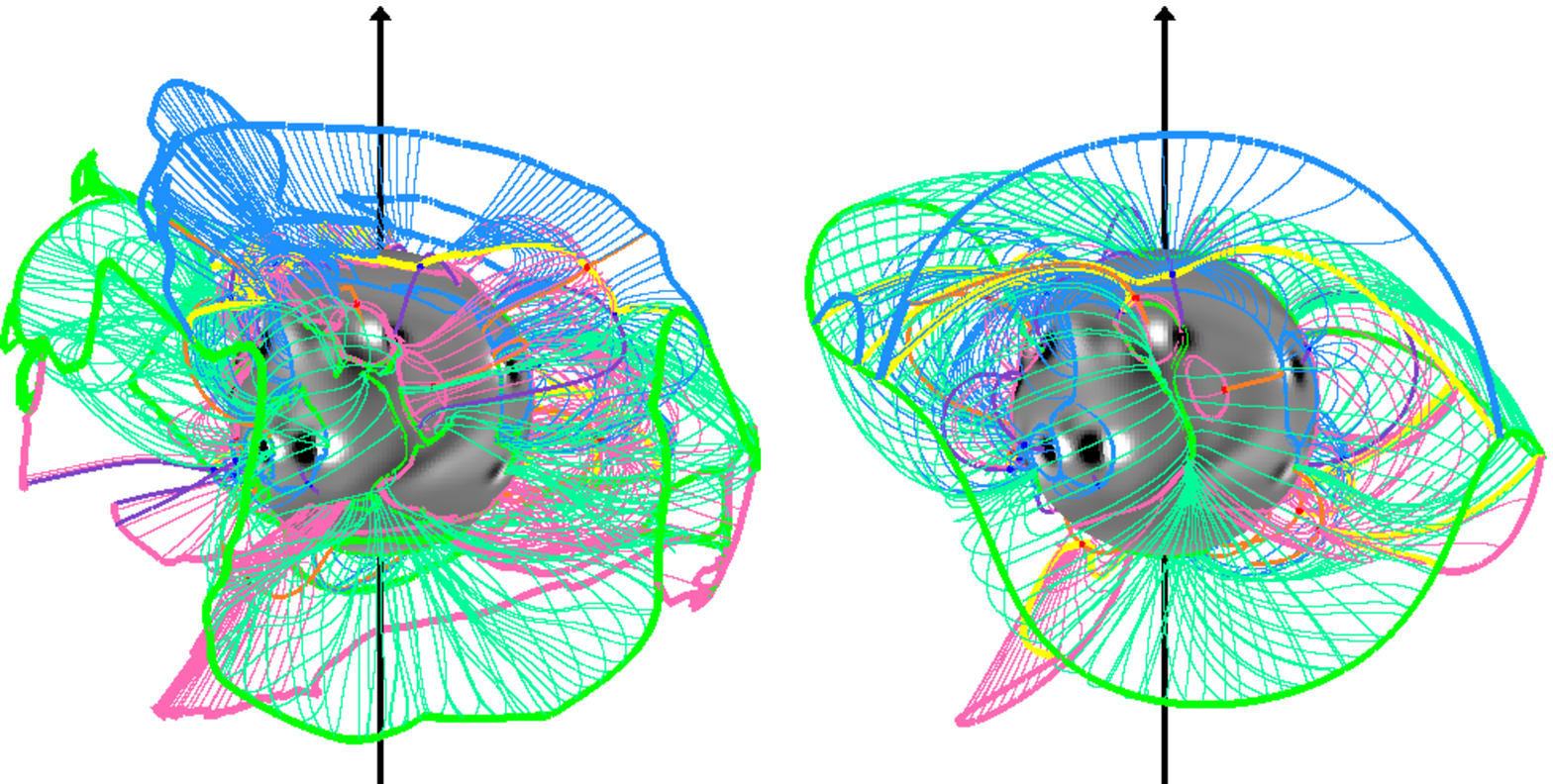}
\caption{Three-dimensional representation of the magnetic skeleton between $1\,\mathrm{R}_\odot$ and $2.5\,\mathrm{R}_\odot$ for 30 April 2013. Left shows the NP model and right the PFSS model. Colours and symbols are as in Figure \ref{fig:skeleton3D_00100}. (Animated versions of these figures are available in the Electronic Supplementary Materials.)}
\label{fig:skeleton3D_00441}
\end{figure}

Figure \ref{fig:skeleton3D_00100} shows the magnetic skeletons for the solar corona between $1\,\mathrm{R}_\odot$ and $2.5\,\mathrm{R}_\odot$ for the NP (left) and PFSS (right) models for 4 April 2000. Correspondingly, Figure \ref{fig:sepsurf_cuts_00100} shows the intersections of the separatrix surfaces with the spherical boundary surface at $2.5\,\mathrm{R}_\odot$. We see that in both models the heliospheric current sheet (hereafter HCS; thick-green line) is very warped, and in the case of the NP field (left) the HCS has split into two disjoint loops as is typical for solar maximum \citep[\textit{e.g.}][]{wang14,platten14}. 
Otherwise, the large-scale structure is broadly similar in the two models except that the PFSS field is much smoother than the NP field. There are, however, more separators and separatrix curtains (separatrix surfaces that reach the source surface) in the NP field. The twisting of the magnetic field when currents are present can cause more separators to form and multiple separators connecting the same pairs of nulls \citep[\textit{e.g.}][]{parnell10}. The greater number of separatrix curtains is associated with the greater amount of open field in the NP case. Again, this is due to the presence of currents, which cause the structures to expand, possibly past the $2.5\,\mathrm{R}_\odot$ boundary \citep{yeates10}. When structures expand close to this boundary, they tend to become open due to the outflow boundary condition imposed in the NP model (which mimicks the effect of the real solar wind). The additional open field is evident in the field-line plots in Figure \ref{fig:3D_fieldlines} and in the maps of the coronal holes in Figure \ref{fig:coronal_holes}.

Figure \ref{fig:sepsurf_cuts_00100} also shows the separatrix surfaces mapped out to $21.5\,\mathrm{R}_\odot$ for the NP and PFSS models. Although this results in a topologically equivalent pattern, the separatrices map to different latitudes and/or longitudes. For example, the region of open field inside the detached HCS loop (in the NP model) occupies a smaller proportion of the spherical surface at $21.5\,\mathrm{R}_\odot$ than at $2.5\,\mathrm{R}_\odot$, implying that field in this region is expanding sub-radially.

\begin{figure}
\includegraphics[width=\linewidth]{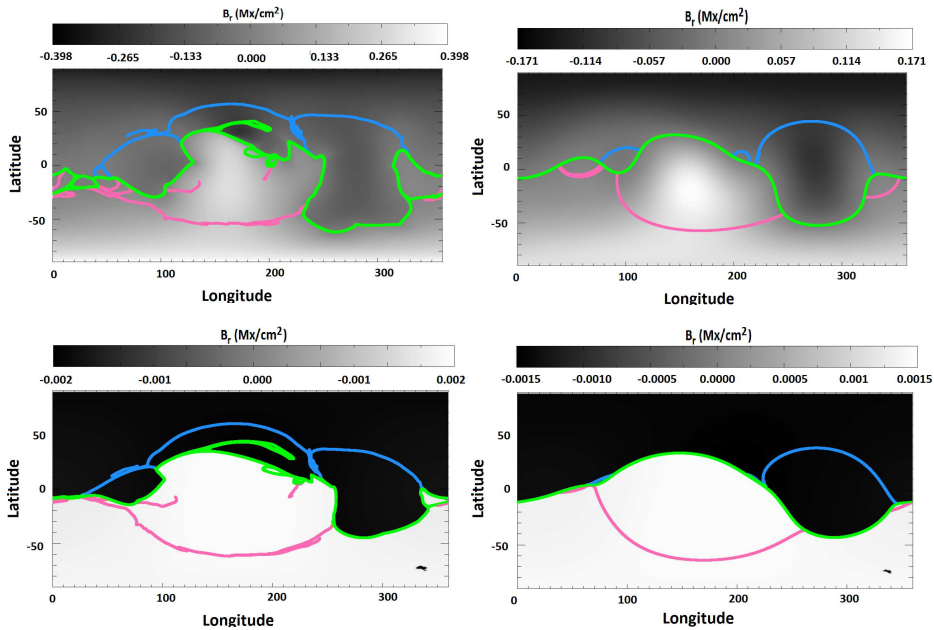}
\caption{Cuts through the magnetic skeleton at $2.5\,\mathrm{R}_\odot$ (top) and $21.5\,\mathrm{R}_\odot$ (bottom) for the NP model (left) and PFSS model (right) for 30 April 2013. The green lines represent the intersection of the HCS with the surface; the pink and blue lines represent the intersection of separatrix surfaces from positive and negative nulls, respectively, with the surface. The background is shaded with $B_r$ (white positive, black negative).}
\label{fig:sepsurf_cuts_00441}
\end{figure}

Similarly, in Figures \ref{fig:skeleton3D_00441} and \ref{fig:sepsurf_cuts_00441} we can compare the magnetic skeletons for 30 April 2013. Here it is clear that the PFSS model is again much smoother, both in the shape of the HCS and also in the shapes and positions of the separatrix curtains. The positive and negative separatrix surfaces in the northern hemisphere interact much more in the NP model, creating many more and also longer separators than in the PFSS model. The HCS itself is much more distorted in the NP model, crossing some longitudes many times. The magnetic field in these folds then contracts sub-radially out to $21.5\,\mathrm{R}_\odot$. Several of the open field regions also contract sub-radially in the PFSS model, with some becoming too small to distinguish at $21.5\,\mathrm{R}_\odot$.

\begin{figure}
\centering
\includegraphics[width=\linewidth]{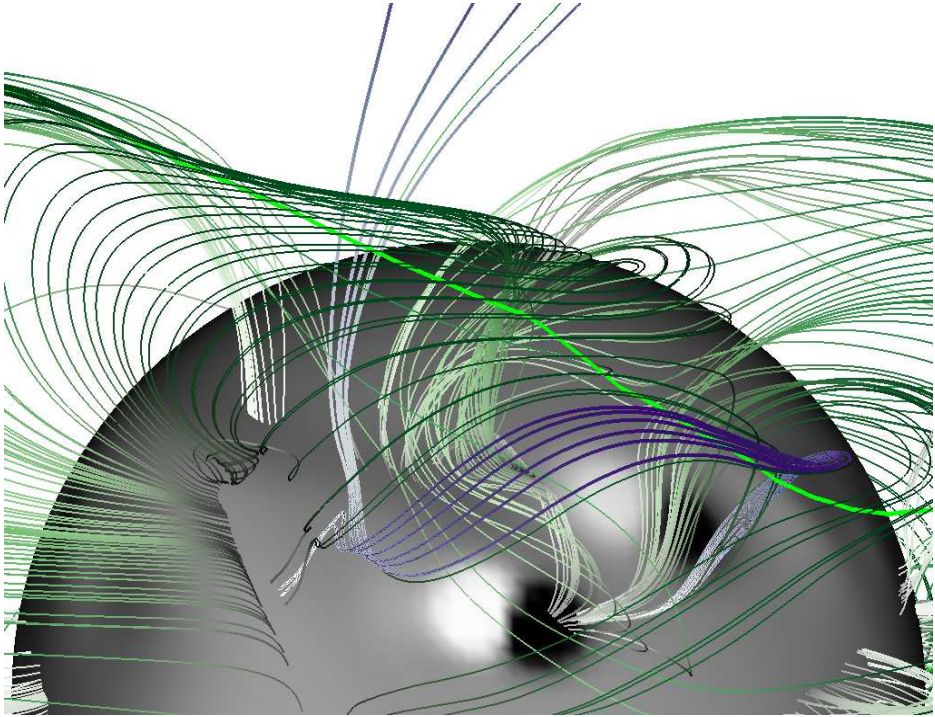}
\caption{Field lines traced from HCS in the NP model for 30th April 2013. Field lines that map down to the photosphere are coloured green. Field lines in the HCS curtains that wrap under the flux rope and map back up to the outer boundary are coloured purple. Field lines are shaded with height above the photosphere with white being closest to the solar surface.}
\label{fig:fieldlines_00441}
\end{figure}

A notable difference between the NP and PFSS models is in the magnetic surfaces traced down from the HCS. In the PFSS model, field lines in the surfaces traced from the HCS all map down to the photosphere and divide open-field regions from closed-field regions. In the NP model this is not always the case. If, for instance, a flux rope is sitting below the base of the HCS and aligned along it, then field lines traced down from the HCS can pass below the flux rope and loop back up to the $2.5\,\mathrm{R}_\odot$ boundary. Such a structure can clearly be seen in the example from 30 April 2013. A 3D image showing example field lines of this type is shown in Figure \ref{fig:fieldlines_00441} (field lines highlighted in purple are ``U-shaped"). In Figure \ref{fig:sepsurf_cuts_00441} (left), these ``U-shaped'' field lines create an apparent excursion of the green HCS curve into the negative polarity region. This excursion is not a polarity-inversion line but merely indicates the opposite end of these field lines traced from the true HCS.

\begin{figure}
\centering
\includegraphics[width=\linewidth]{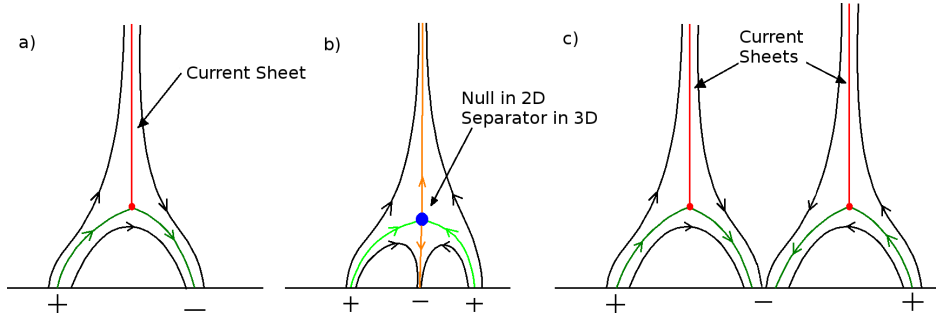}
\caption{Sketch of the cross section of a streamer (a), a pseudostreamer (b) and a double streamer (c).}
\label{fig:streamer_sketch}
\end{figure}

Observationally, the HCS and associated separatrix surfaces are identified as streamers or helmet streamers. Separatrix curtains from null points also form streamer-like structures. These structures differ from helmet streamers because there is no polarity change across the streamer and so they are known as pseudostreamers. The difference is illustrated in Figures \ref{fig:streamer_sketch}a and b. In Figures \ref{fig:skeleton3D_00100} to \ref{fig:sepsurf_cuts_00441}, the green magnetic surfaces traced down from the HCS correspond to streamers, whereas the separatrix curtains (pink and blue surfaces) cutting the outer boundary correspond to pseudostreamers. It should be noted that not all of these streamer structures would be observed in coronagraphs as the streamer needs to be aligned along the line of sight in order to be visible.

The radial inflation of the field in the NP model can lead to structures that appeared as pseudostreamers in the PFSS model becoming double streamers in the NP model (see Figure \ref{fig:streamer_sketch}b and c). Effectively the null point has risen out of the domain. A good example of this is seen in Figure \ref{fig:sepsurf_cuts_00100}: where the HCS has split into two loops in the NP model, there is a separatrix surface in the PFSS model. We note that it is possible to have double streamers also in the PFSS model, as would occur if the central photospheric polarity is strong enough \citep[][]{rachmeler14}.

\subsection{Squashing Factor}

As well as examining the magnetic skeleton we also examine the ``quasi-skeleton''. We trace field lines down from the $21.5\,\mathrm{R}_\odot$ outer boundary to the photosphere, $1\,\mathrm{R}_\odot$, and calculate the squashing factor [$Q$] of this mapping, using the definition of \citet{titov07} appropriate for spherical coordinates:
\begin{equation}
Q = N^2/|\Delta|
\end{equation}
where
\begin{equation}
N^2=\frac{R_*^2}{R^2}\left[\left(\frac{\sin(\Theta)}{\sin(\theta)}\frac{\partial \Phi}{\partial \phi}\right)^2+\left(\sin(\Theta)\frac{\partial \Phi}{\partial \theta}\right)^2+\left(\frac{1}{\sin(\theta)}\frac{\partial \Theta}{\partial \phi} \right)^2+\left(\frac{\partial \Theta}{\partial \theta}\right)^2\right],
\end{equation}
and 
\begin{equation}
\Delta=B_r/B^*_r,
\end{equation}
where $(\Theta(\theta,\phi),\Phi(\theta,\phi))$ is the field line mapping from a sphere of radius $R$ to one of radius $R_*$ and $B_r$ and $B^*_r$ are the components of the magnetic field normal to the boundary at each end of the field line.

This is a measure of gradients in the field-line mapping; so-called quasi-separatrix layers (QSLs) of high $Q$ are locations where nearby magnetic-field lines diverge strongly.
\begin{figure}
\includegraphics[width=\linewidth]{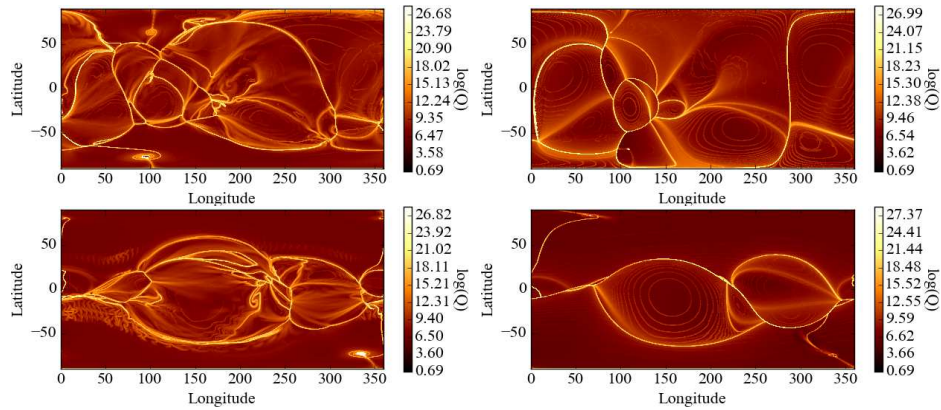}
\caption{Squashing factor [$Q$] calculated at $21.5\,\mathrm{R}_\odot$ boundary for 4 April 2000 (top row) and 30 April 2013 (bottom row) for the NP (left column) and PFSS (right column) models.}
\label{fig:Q}
\end{figure}
Figure \ref{fig:Q} shows the squashing factor on this upper boundary. We see that many of the structures match the locations of the separatrix-surface cuts through the $21.5\,\mathrm{R}_\odot$ boundary shown in Figure \ref{fig:sepsurf_cuts_00100} and Figure \ref{fig:sepsurf_cuts_00441} (bottom rows). These structures are true discontinuities in the field line mapping. However, we see extra features in $Q$ that do not appear in the skeleton. While separatrix curtains in the skeleton separate disconnected coronal holes at the photosphere, the additional QSLs seen in the map of $Q$ represent locations where open field lines undergo rapid but continuous changes in footpoint location from one part of the photosphere to another. They typically divide open-field regions of the same polarity between different photospheric coronal holes that are connected by narrow corridors of open field (since the mapping is continuous). This is the basis of the ``S-web'' model of \citet{antiochos11}, who suggest that the presence of these narrow open-field corridors can explain why slow solar wind is found across a wide angular range on the Sun, not just at the boundaries of polar coronal holes. In this scenario, our comparison indicates that the NP model provides more source regions for the slow solar wind than the PFSS model.

We note that the spiral features of high $Q$ seen near the South Pole at approximately 100$^\circ$ longitude on 4 April 2000 and approximately 340$^\circ$ longitude on 30 April 2013 are associated with field lines twisting around the poles as the Sun rotates; the simulation domain of the NP model extends to only $89.5^\circ$ so the mapping cannot be continuous across the poles.

\section{Effect on Empirical Wind Speeds}\label{s:proxies}

In the WSA model, radial wind speed at $21.5\,\mathrm{R}_\odot$ is determined by an empirical formula based solely on the modeled magnetic-field structure. For example, one current implementation (based on GONG and NSO magnetograms) uses the formula (C.N. Arge, private communication)
\begin{equation}
v_r(\theta,\phi)=240.0+\frac{675}{(1+f_\mathrm{s})^{1/4.5}}\left[1.0-0.8e^{(-(\theta_\mathrm{b}/1.9)^{2})}\right]^{3} \mathrm{km}\,\mathrm{s}^{-1},
\label{eqn:vr}
\end{equation}
where $f_\mathrm{s}(\theta,\phi)$ measures the local radial expansion of a magnetic flux tube, and $\theta_\mathrm{b}(\theta,\phi)$ is the angular distance of the field line's photospheric footpoint from the nearest coronal-hole boundary. This is a slight modification of the formula discussed by \citet{arge04}. Here we study how the factors $f_\mathrm{s}$ and $\theta_\mathrm{b}$ differ between the NP and PFSS models. For completeness, we show the corresponding $v_r$ distributions computed with Equation (\ref{eqn:vr}), but we caution that the numerical factors in this formula have been optimised to best match solar-wind observations for the PFSS model. It is known that these empirical wind models are sensitive to the type of magnetic-field model used \citep[\textit{e.g.}][]{riley15}. Optimizing such an empirical formula for the NP model will be carried out in the future.

\subsection{Flux Tube Expansion}

\begin{figure}
\centering
\includegraphics[width=\linewidth]{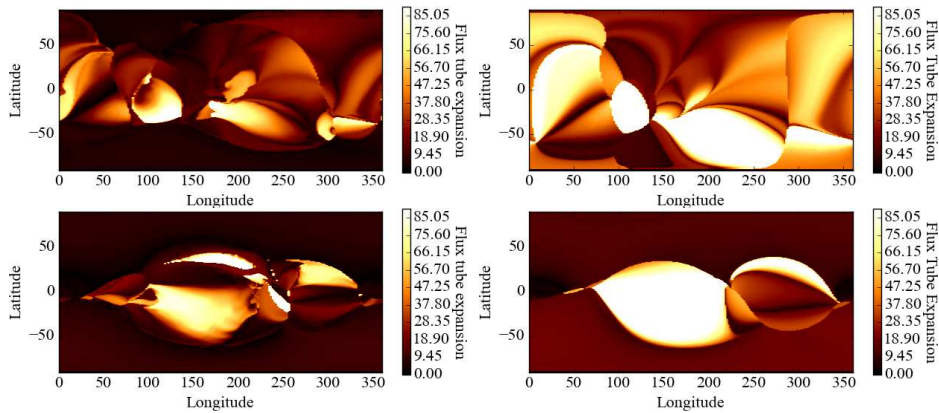}
\caption{Flux-tube expansion [$f_\mathrm{s}$] at $21.5\,\mathrm{R}_\odot$ for 4 April 2000 (top) and 30 April 2013 (bottom) for the NP model (left) and the PFSS model (right).}
\label{fig:fte}
\end{figure}

Figure \ref{fig:fte} shows the flux-tube expansion [$f_\mathrm{s}$] calculated between $1\,\mathrm{R}_\odot$ and $2.5\,\mathrm{R}_\odot$ and then mapped along the field lines to $21.5\,\mathrm{R}_\odot$ for each of the cases. This quantity was introduced by \citet{wang90} as a proxy for the solar-wind speed in purely magnetic models. Greater flux-tube expansion was shown to correlate with slower wind speed. The flux-tube expansion is defined as the ratio of the magnetic-field components at each end of the field line normal to the two boundaries,
\begin{equation}
f_\mathrm{s}=\left(\frac{\mathrm{R}_\odot}{\mathrm{R}_\mathrm{s}}\right)^2\left(\frac{B_r(\mathrm{R}_\odot)}{B_r(\mathrm{R}_\mathrm{s})}\right)
\end{equation}
where, in our case, $\mathrm{R}_\mathrm{s}=2.5\,\mathrm{R}_\odot$.

We see that in all cases the pattern of $f_\mathrm{s}$ on the $21.5\,\mathrm{R}_\odot$ surface is structured by the separatrix-surface cuts (Figures \ref{fig:sepsurf_cuts_00100} and \ref{fig:sepsurf_cuts_00441}), but clearly also reflects the variations in field-line divergence shown by the squashing factor [$Q$] (Figure \ref{fig:Q}). Overall there is a greater expansion in the PFSS models (right-hand column) than in the corresponding NP models (left-hand column). This is because the photospheric open-field regions that correspond to the base of coronal holes are much smaller in the PFSS model (see Figure \ref{fig:coronal_holes}). Since all field must be open at $2.5\,\mathrm{R}_\odot$, smaller photospheric open-field regions will imply greater expansion factors.

\subsection{Coronal Hole Boundary Distance}

\begin{figure}
\centering
\includegraphics[width=\linewidth]{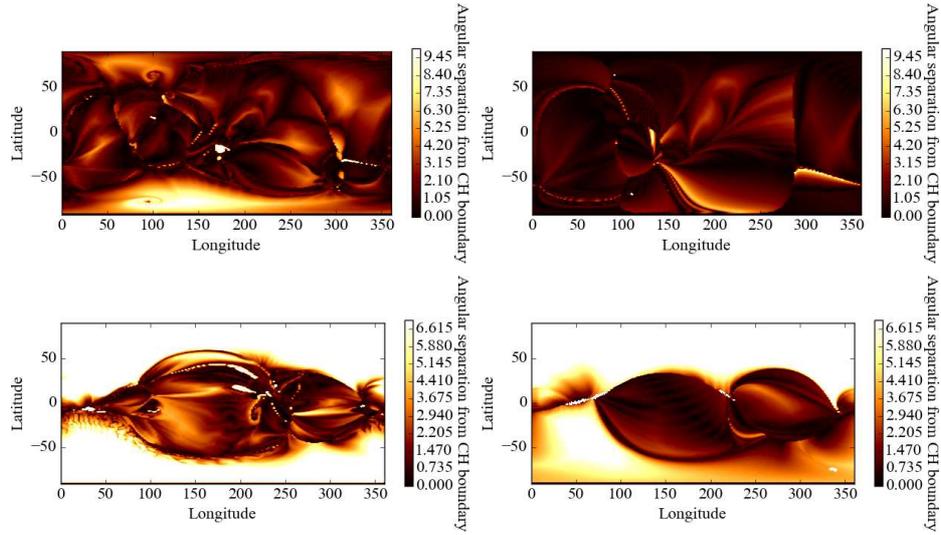}
\caption{Distance [$\theta_\mathrm{b}$] from coronal hole boundary of footpoints of fieldlines traced from $21.5\,\mathrm{R}_\odot$ for 4 April 2000 (top) and 30 April 2013 (bottom) for the NP model (left) and the PFSS model (right).}
\label{fig:chd}
\end{figure}

\citet{riley01} proposed a relationship between the wind speed on an open magnetic field-line and the distance of that field line's footpoint from the nearest coronal-hole boundary. It was incorporated into the WSA model by \citet{arge03} to account for discrepancies in wind speeds along field lines that had the same flux-tube expansion factor. In fact, it is a much more important factor in determining the wind speed in Equation (\ref{eqn:vr}) than the flux-tube expansion \citep{riley15}. We measure it as the angular separation of the photospheric footpoints from their nearest coronal-hole boundary. We describe the boundaries of the coronal holes (as shown in Figure \ref{fig:coronal_holes}) using a series of points, and for each open-field-line footpoint we calculate the great-circle distance to the nearest coronal-hole boundary point. From this we then calculate the angular separation. Figure \ref{fig:chd} shows maps of this quantity for field lines traced down from $21.5\,\mathrm{R}_\odot$ to $1\,\mathrm{R}_\odot$. The largest distances from the edges of coronal holes are seen on 30 April 2013 in the NP model (bottom left Figure \ref{fig:chd}) since this frame contains the largest coronal holes, as seen from Figure \ref{fig:coronal_holes} (bottom-left). 

\subsection{Empirical Wind Speed}

\begin{figure}
\centering
\includegraphics[width=\linewidth]{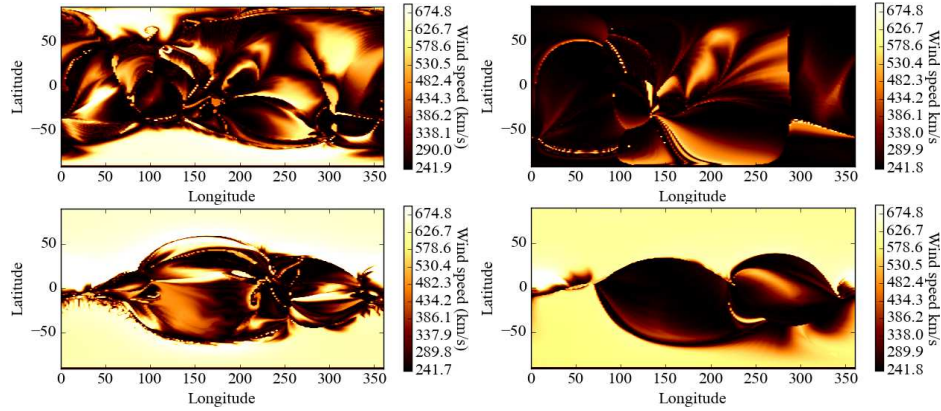}
\caption{Distribution [$v_r(\theta,\phi)$] of empirical solar wind speed at $21.5\,\mathrm{R}_\odot$ for 4 April 2000 (top) and 30 April 2013 (bottom) for the NP model (left) and the PFSS model (right).}
\label{fig:vr}
\end{figure}

\begin{figure}
\includegraphics[width=\linewidth]{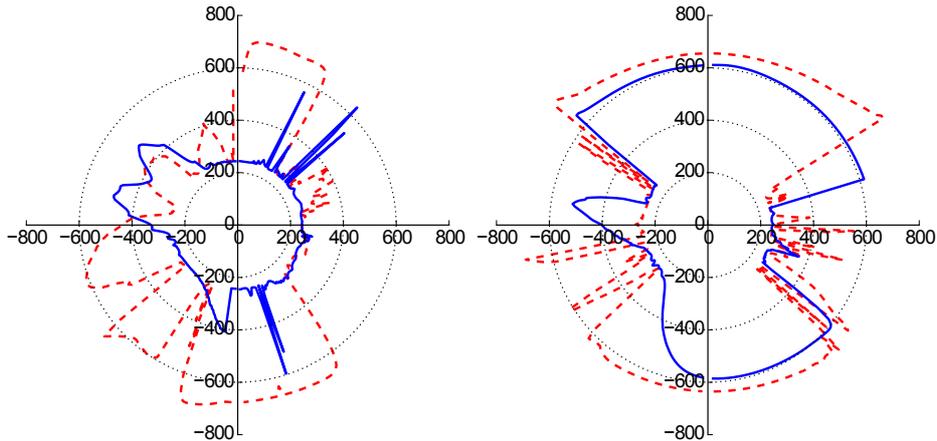}
\caption{Empirical solar-wind speed in the plane of sky viewed from Carrington longitude $0^\circ$ on 4 April 2000 (left) and 30 April 2013 (right). Red-dashed shows NP model, Blue-solid shows PFSS model.}
\label{fig:ulyssestype}
\end{figure}

Figure \ref{fig:vr} shows the empirical solar-wind speed at $21.5\,\mathrm{R}_\odot$ calculated using Equation (\ref{eqn:vr}). The highest wind speeds occur in the NP field model and out of the two examples given the wind was faster on 30 April 2013. The wind speed calculated using the potential field for 4 April 2000 shows the overall slower wind speed. Visual comparison of Figure \ref{fig:vr} with Figures \ref{fig:fte} and \ref{fig:chd} shows that the distribution of empirical wind speed derives primarily from the coronal-hole boundary distance [$\theta_\mathrm{b}$], rather than the flux-tube expansion [$f_\mathrm{s}$]. This is a result of the formula in Equation (\ref{eqn:vr}), and is in accordance with \citet{riley15}.

Figure \ref{fig:ulyssestype} shows a polar plot of the empirical solar-wind speed in the plane of sky, from one particular viewing angle. The speeds for the NP (red) and PFSS (blue) models are overlaid. The additional solar-wind structure predicted by the NP model at lower latitudes is evident, particularly on 30 April 2013.

The highest wind speeds correspond to locations where there are large coronal holes. The southern polar coronal hole shows speeds of over 650$\,{\rm km}\,{\rm s}^{-1}$ in the NP model on 4 April 2000, but this coronal hole is not present in the PFSS model for the same date. This is clearly visible in Figure \ref{fig:ulyssestype} (left) where we see fast wind at both poles in the NP model but not in the PFSS model. On 30 April 2013, both models have substantial polar coronal holes, but the wind speeds in the NP model are higher owing to the reduced horizontal expansion of the coronal holes with height, compared with the PFSS model (cf. Figure \ref{fig:3D_fieldlines}).

At lower latitudes, Figure \ref{fig:vr} clearly shows additional sub-structure in the NP wind speed as compared to the PFSS wind speed. Such differences would be observed at 1 AU as additional temporal fluctuations in the predicted wind speed, as compared to current WSA forecasts. The difference reflects both the larger number of low-latitude coronal holes in the NP model, but also the more complex magnetic structure of the NP model. The latter was demonstrated by the more complex skeleton and quasi-skeleton found in Section \ref{s:top}.

\section{Conclusions}
In summary, we have shown that removing the potential field (current-free) assumption used in solar-wind models such as WSA could have an important impact on predicted wind-speed distributions. In comparing magneto-frictional non-potential (NP) and potential field source surface (PFSS) models on two dates near to solar maximum, we have found significant differences in the latitude--longitude distribution of predicted solar wind speed at 0.1 AU. If extrapolated using, for example, the Enlil model, these differences would in turn lead to significant differences in temporal variations of predicted wind speed at 1 AU. Since we used identical photospheric boundary conditions (magnetic maps) in both models, our results suggest that the uncertainty due to omission of coronal electric currents in existing models is likely to be at least as large as that due to the use of magnetogram data from different observatories \citep[cf.][]{riley14}. To give one example, we found substantial polar coronal holes in the NP model for 4 April 2000, despite their absence in the PFSS model.

Although the formula we use for empirical wind speed (Equation (\ref{eqn:vr})) has been optimized specifically for the PFSS model, the differences between the predicted speed in the NP and PFSS models arise from differences in the basic physical quantities used in the formula, namely, the expansion rate of open magnetic flux tubes and the distance of their footpoints from coronal-hole boundaries on the solar photosphere. Ultimately the cause is a difference in the topological structure of the coronal magnetic field. The key difference is that ``inflation'' of the coronal field by the presence of electric currents in the NP model leads to additional open magnetic field and more coronal holes (open-field regions). This in turn leads to lower flux-tube expansion factors than in the PFSS model, at the same time as shorter footpoint distances to coronal hole boundaries, particularly at lower latitudes. 

While specific details will vary (and will in general be quite sensitive to model input and parameters), these general conclusions are not specific to the NP model. We would expect to see similar differences in full MHD models where non-potential magnetic structure is built up through continuous driving. Indeed, \citet{riley06} already found more open field in their MHD model compared to a PFSS model, although the difference in that case was less pronounced because the MHD model was initiated from a PFSS extrapolation and relaxed to equilibrium, rather than driving the photospheric field continuously over a longer period. When continuous driving of the photospheric field is included, the NP model suggests that more complex topologies may form in the magnetic field, such as twisted structures. We have seen how the HCS separatrix surfaces no longer always separate closed field from open field. Moreover, in the presence of coronal currents, open field lines may no longer have an anchoring point at the photosphere (cf. the U-shaped field lines in Figure \ref{fig:fieldlines_00441}). All of these differences will have an impact on solar-wind speed predictions.

In the future, we plan to couple output from this model to the Enlil model, in order that we can test the model against time series of in-situ wind speed (and polarity) measurements at 1 AU. This will require us to optimize the empirical formula for wind speed to the NP model; in this article, we used the existing PFSS-based formula to study the differences between the PFSS and NP models. An important uncertainty to quantify will be the sensitivity of the predicted wind speeds to differences in the photospheric input.

%%%%%%%%%%%%%%%%%%%%%%%%%%%%%%%%%%%%%%%%%%%%%%%%%%%%%%%%%%%%%%%%%%%%%%%%%%%
%% Appendix
%
% \appendix   

%%%%%%%%%%%%%%%%%%%%%%%%%%%%%%%%%%%%%%%%%%%%%%%%%%%%%%%%%%%%%%%%%%%%%%%%%%%
%% Acknowledgements
%
 \begin{acks}
A.R. Yeates and S.J. Edwards were supported by STFC through consortium grant ST/K001043/1 and the Durham University Impact Acceleration Account, as well as by the US Air Force Office for Scientific Research. D.H. Mackay would like to thank the Leverhulme Trust and STFC for financial support. The authors thank Andrew L. Haynes for the use of his separatrix surface and null-point finding codes. Numerical simulations used the SRIF and STFC funded HPC cluster at the University of St Andrews.
 \end{acks}

\section*{Disclosure of Potential Conflicts of Interest}
The authors declare that they have no conflicts of interest.

%%% %%%%%%%%%%%%%%%%%%%%%%%%%%%%%%%%%%%%%%%%%%%%%%%%%%%%%%%%%%%
%% Bibliography
%
% Using BibTeX
%
 \bibliographystyle{spr-mp-sola}
% %\bibliographystyle{spr-mp-sola-limited} %% Alternative style: no title
 \bibliography{bibliography.bib}  
%
% Without BibTeX 
% \begin{thebibliography}{}
% \bibitem[\protect\citeauthoryear{Author}{Year}]{key}
%   <bibliographical entry>
%
% \bibitem[\protect\citeauthoryear{}{}]{}
%   
%  
% \end{thebibliography}

\end{article} 
\end{document}